\documentclass[aps,prl,twocolumn,
superscriptaddress,showpacs,amsmath,amssymb,unsortedaddress]{revtex4}
\usepackage{graphics}
\usepackage{epsfig}
\usepackage{color}
\begin{document}
%
%
\title{Anomalous Light Absorption by Small Particles}
%
%
\author{Michael I. Tribelsky}
\email[]{tribelsky_at_mirea.ru}
\affiliation{Moscow State Institute of Radioengineering, Electronics
and Automation (Technical University), 78 Vernadskiy Ave., Moscow
119454, Russia}
\affiliation{Max-Planck-Institut f\"ur Physik komplexer Systeme,
N\"othnitzer Str. 38, Dresden 01187, Germany}

\date{\today}
%
%
\begin{abstract}
 A new type of resonant light absorption by a small particle (nanocluster) is reported. The problem cannot be described within the commonly used dipole scattering approximation and should be studied with methods based upon the exact Mie solution. It is shown that the absorption cross-section has giant maxima realized at \emph{small\/} values of the imaginary part of the complex dielectric permittivity of the particle. The maxima are situated in the vicinity of the plasmon (polariton) resonances and correspond to the regions where the dissipative damping equals the radiative one. The case is similar to the recently introduced anomalous scattering [PRL \textbf{97}, 263902 (2006)] and exhibits similar peculiarities.
\end{abstract}


\maketitle
%
%
Since the first quantitative study by Lord Rayleigh in 1871~\cite{Rayleigh}, the problem of light scattering by small particles has remained one of the most important and appealing issues of electrodynamics. There are thousands of articles and numerous monographs devoted to this matter, see e.g.~\cite{Book} and references therein. Nonetheless, the problem is far from completion.

Recently it was pointed out~\cite{MIT-Luk,MIT-Fano,OPN} that small particles made of weakly dissipating materials may exhibit the so-called \textit{anomalous scattering} (AS), which has very little in common with Rayleigh scattering, other than both occur without changes in the frequency of incident light $\omega$. AS results in giant optical resonances with narrow linewidths and inverted hierarchy, i.e., the partial extinction cross-section for the quadrupole resonance is \emph{larger} than that for the dipole one, etc.~\cite{MIT_JETP,n1}.  It occurs when the dissipative losses are small enough to become negligible relative to the radiative damping~\cite{MIT-Luk,MIT-Fano,OPN}. Therefore, it seems that the phenomenon has nothing to do with the maximal absorption of the light by the particle, when the dissipation should be as large as possible. However, this is not the case.

In this Letter, the absorption cross-section $\sigma_{abs}(R, \epsilon'_p, \epsilon''_p)$ of a spherical particle with radius $R$ and complex dielectric permittivity $\epsilon_p = \epsilon'_p + i\epsilon''_p$ is analyzed, based upon the exact Mie solution~\cite{Mie}. The exact solution automatically includes both types of damping (dissipative and radiative). Thus, the problem is just to find an appropriate asymptotic of the cumbersome Mie solution. Such an analysis indicates that $\sigma_{abs}(R, \epsilon'_p, \epsilon''_p)$ for small particles exhibits sharp giant peaks whose maxima lie at $\epsilon_p'' \ll 1$, while the corresponding values of $\epsilon_p'$ coincide with those for plasmon (polariton) resonances. It should be stressed that apart from purely academic interest, the problem of maximization of $\sigma_{abs}$ is important for a number of applications of nanoparticles in technology, biology, medicine, etc. -- which provides additional stimuli for the present study.

To understand the physical nature of the aforementioned peculiarities of $\sigma_{abs}$, let us estimate the power $P$ dissipated in a unit volume of the particle. Neglecting the magnetic part of the dissipation owing to the particle smallness~\cite{LL}, one obtains $P \propto \epsilon''|\mathbf{E}|^2$, where $\mathbf{E(r)}$ is a complex amplitude of the monochromatic electric field in a given point inside the particle. On the other hand, in the vicinity of each plasmon (polariton) resonance $\mathbf{E(r)}$ undergoes a sharp increase. At these resonances in the case of Rayleigh scattering, $\mathbf{E(r)}$  is cut off by dissipative losses, so that $\mathbf{E(r)_\ell} \propto 1/\epsilon''_p$. Thus at the resonances, $P \propto 1/\epsilon''_p \rightarrow \infty$ at $\epsilon''_p \rightarrow 0$, which gives rise to the corresponding divergence of $\sigma_{abs}$.

The divergence occurs because the approximation used is invalid. In fact, $\sigma_{abs}$ does not diverge. At small enough $\epsilon''_p$, the radiative damping prevails over the dissipative losses and Rayleigh scattering is replaced by AS, which provides a different cutoff for $\mathbf{E(r)}$~\cite{MIT-Luk,MIT-Fano,OPN,MIT_JETP}.

To inspect the problem accurately, one should employ the exact Mie solution. It is convenient to introduce dimensionless cross-sections $Q_{ext, sca, abs} = \sigma_{ext, sca, abs}/\pi R^2,$ where $\sigma_{ext}, \sigma_{abs}$ and $\sigma_{sca}$ respectively stand for the extinction,  scattering  and absorption cross-sections of the particle. Then, the Mie solution yields the following well-known expressions:
\begin{equation}\label{sigma}
Q_{ext} \!\! \simeq \!\! \sum\limits_{l \!
=\! 1}^\infty {Q_{ext}^{(l )} },\:Q_{sca}  =
\sum\limits_{l = 1}^\infty  {Q_{sca}^{(l )} },\: Q_{abs}\! = \!Q_{ext} \! - \!Q_{sca} ;
\end{equation}
\begin{equation}\label{sigmale}
 Q_{ext}^{(l )}  = \frac{{2}}{{q^2 }}\left( {2l + 1}
\right){\mathop{\rm Re}\nolimits} \left( {a_l   +
b_l } \right),
\end{equation}
\begin{equation}\label{sigmals}
 Q_{sca}^{(l )}  = \frac{{2}}{{q^2}}\left( {2l + 1}
\right)\left( {|a_l  |^2  + |b_l  |^2 } \right).
\end{equation}
%
%
The summation is over the corresponding partial multipole cross-sections, the so-called size parameter $q$ equals $n_m Rk_0,\; n_m = \sqrt \epsilon_m$ is purely real refractive index of the environmental medium, and $k_0$ stands for the wavenumber of the incident light in vacuum ($k_0 = \omega/c$, where $c$ is the speed of light).
%

Amplitudes $a_l$ and $b_l$ may be presented as follows~\cite{MIT-Luk}:

\begin{equation}
\label{a} a_l  = \frac{F^{(a)}_l  \left( {q ,\epsilon}
\right)}{F^{(a)}_l  \left( {q ,\epsilon} \right) + iG^{(a)}_l
\left( {q ,\epsilon} \right)}.
\end{equation}
Here $\epsilon = \epsilon_p(\omega)/\epsilon_m(\omega)$ and
$F^{(a)}_l \rightarrow F^{(b)}_l; \,\,G^{(a)}_l \rightarrow
G^{(b)}_l$ for $b_l$. $F^{(a,b)}_l,\,\, G^{(a,b)}_l$ are
expressed in terms of the Bessel [$J_{l+1/2}(z)$] and Neumann
[$N_{l+1/2}(z)$] functions, whose expansions in power series gives
rise to the following formulae, valid at small $q$:
\begin{eqnarray}
  &  &\hspace*{-8mm}F^{(a)}_l\!\left(q,\epsilon\right) \!
  \simeq \!q^{2l  + 1}\frac{l  + 1}{{\left[
{\left( {2l  + 1} \right)!!} \right]^2 }}\left( {\epsilon -
1} \right)+... \label{Raq}\\
  &  &G^{(a)}_l(q,\epsilon)  \simeq
  \frac{l }{2l  + 1}\left\{\epsilon +
\frac{l  + 1}{l }  -\right. \nonumber \\
     & &\left. q^2\frac{\epsilon - 1}{2}
     \left[\frac{\epsilon}{2l  + 3} + \frac{l  +
1}{l \left(2l  - 1\right)} \right] +... \right\}, \label{Iaq}
  \end{eqnarray}
(ellipses denote omitted higher order terms in $q$). The explicit expressions for $F^{(b)}$ and $G^{(b)}$ are not required because of the estimate $|b_l| \sim
q^{2l+3} \ll |a_l|$, which allows neglect of $b_l$ relative to $a_l$ at $q \ll 1$.

Eq.~(\ref{a}) results in the following expression for the partial absorption cross-sections $Q^{(l)}_{abs} = Q^{(l)}_{ext} - Q^{(l)}_{sca}$:
\begin{equation}\label{Ql}
    Q^{(l)}_{abs}\! = \!\frac{2(2l+1)}{q^2}\frac{{F^{(a)}_l}'' {G^{(a)}_l}' - {F^{(a)}_l}' {G^{(a)}_l}'' }{({F^{(a)}_l}'\!\! - {G^{(a)}_l}'')^2 + ({F^{(a)}_l}''\!\! + {G^{(a)}_l}')^2 },
\end{equation}
where prime and double prime denote the corresponding real and imaginary parts, respectively.

The plasmon (polariton) resonances are defined by the condition ${G^{(a)}}'_l(q,\epsilon') = 0$, which determines the resonant values of the real part of the dielectric permittivity $\epsilon_l'(q)$ and through the dependence $\epsilon(\omega)$, the resonant frequencies $\omega_l(q)$. For the problem in question the maxima of $\sigma_{abs}$ should be situated in the vicinity of the resonances. Let us inspect the vicinity of a certain $\ell$th resonance, presenting the dielectric permittivity in the form
\begin{equation}\label{delta_epsilon}
    \epsilon = -\frac{\ell+1}{\ell} + \delta\epsilon' + i\epsilon'',
\end{equation}
where $\delta\epsilon'$ and $\epsilon''$ are small, see Eq.~(\ref{Iaq}). Then, the leading approximations for $F^{(a)}_\ell$, $G^{(a)}_\ell$ and $Q^{(\ell)}_{abs}$ read:
\begin{eqnarray}
  F^{(a)}_\ell\!\!\! &\simeq& q^{2\ell+1}\frac{\ell+1}{[(2\ell+1)!!]^2}\left(-\frac{2\ell+1}{\ell}+i\epsilon''\right), \label{Fappr} \\
  G^{(a)}_\ell \!\!\!&\simeq&\!\!\! \frac{\ell}{2\ell+1}\left\{\!\left[\delta\epsilon'\!+\!q^2\frac{4(2\ell\!+\!1)(\ell\!+\!1)}{\ell^2(2\ell\!-\!1)(2\ell\!+\!3)}\right]\! +\! i\epsilon''\!\right\}, \label{Gappr}\\
  Q^{(\ell)}_{abs} \!\!\!&\simeq& \!\! \frac{2(2\ell+1)}{q^2}\frac{-{F^{(a)}_\ell}'{G^{(a)}_\ell}''}{({{G^{(a)}_l}''\!\! - F^{(a)}_l}')^2 + ({G^{(a)}_l}')^2 }. \label{Qappr}
\end{eqnarray}

According to Eq.~(\ref{Fappr}), ${F^{(a)}_\ell}'$ is negative and does not depend on $\delta \epsilon',\; \epsilon''$. It allows one to transfer from $\delta \epsilon',\; \epsilon''$ to ${G^{(a)}_\ell}',\; {G^{(a)}_\ell}''$, considering the latter as new independent variables, see Eq.~(\ref{Gappr}). The maximum of $Q^{(\ell)}_{abs}$ corresponds to ${G^{(a)}_\ell}'= 0$ (i.e., it is achieved exactly at the resonant frequency $\omega_\ell$) and ${G^{(a)}_\ell}''= -{F^{(a)}_\ell}'$ [it should be remembered that ${G^{(a)}_\ell}'' > 0$, see Eq.~(\ref{Gappr})]. Note, the two terms ${G^{(a)}_l}''$ and $-{F^{(a)}_l}'$ in the denominator of Eq.~(\ref{Qappr}) correspond to the dissipative and radiative damping, respectively. So, $Q^{(\ell)}_{abs}$ is maximized when the former equals the latter.

\begin{figure}[h]
\hspace*{-3mm}
  \includegraphics[width=90mm,
  ]{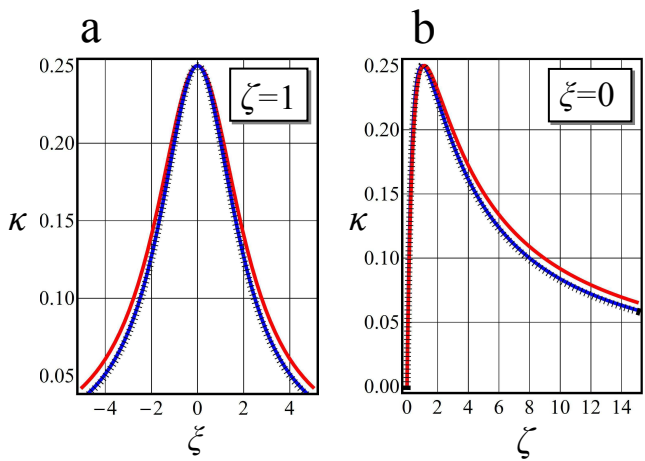}
  \caption{(color online). Resonant absorption lineshapes along axes of real (a) and imaginary (b) parts of dielectric permittivity of the particle, in rescaled dimensionless variables Eq.~(\ref{rescale}). The universal profiles corresponding to the approximate Eq.~(\ref{kappa}) are shown as dotted lines. Lines for the dipole and quadrupole resonances derived from the exact Mie solution for $q=0.3$ are shown as solid red and blue lines, respectively. Note high accuracy of the approximate expression Eq.~(\ref{kappa}); while for the dipole resonance the universal lines given by Eq.~(\ref{kappa}) are still a bit narrower than those for the exact Mie solution, for the quadrupole resonance they are practically identical with the latter.}\label{F1}
\end{figure}

Thus in the vicinity of each plasmon (polariton) resonance, $Q^{(\ell)}_{abs}$ passes through the local maximum
\begin{equation}\label{Qmax}
    Q^{(\ell)}_{abs\;max} = \frac{2\ell+1}{2q^2},
\end{equation}
which equals $1/4$ of the maximal AS resonant partial extinction cross-section, $Q^{(\ell)\; AS}_{ext\;max}$~\cite{MIT_JETP}, and also linearly increases with an increase in the order of resonance $\ell$. However, in contrast to $Q^{(\ell)\; AS}_{ext\;max}$ (achieved at $\epsilon'' = 0$), the maximum of $Q^{(\ell)}_{abs}$ is realized at
\begin{eqnarray}
   \epsilon' &=& \epsilon'_\ell \simeq -\left[\frac{\ell+1}{\ell} + q^2\frac{4(2\ell+1)(\ell+1)}{\ell^2(2\ell-1)(2\ell+3)}\right],\label{Re_e_l}\\
   \epsilon'' &=& \epsilon''_\ell \simeq q^{2\ell+1}\frac{\ell+1}{[\ell(2\ell-1)!!]^2}. \label{Im_e_l}
\end{eqnarray}
The corresponding linewidths measured at the level $Q^{(\ell)}_{abs}=Q^{(\ell)}_{abs\;max}/2$ are as follows:
\begin{eqnarray}
   \Delta \epsilon' &=& q^{2\ell+1}\frac{4(\ell+1)}{[\ell(2\ell-1)!!]^2},\; \epsilon'' = \epsilon''_\ell; \label{Delta_e'}\\
   \Delta \epsilon'' &=& q^{2\ell+1}\frac{4(\ell+1)\sqrt 2}{[\ell(2\ell-1)!!]^2},\; \epsilon' = \epsilon'_\ell. \label{Delta_e''}
\end{eqnarray}
Though the values of $\Delta \epsilon',\;\Delta \epsilon''$ are close to each other, the lineshapes are quite different. To study this issue in detail let us introduce new variables:
\begin{equation}\label{rescale}
    \kappa = \frac {q^2 Q}{2(2\ell+1)},\; \xi = -\frac{{G^{(a)}_\ell}'}{{F^{(a)}_\ell}'},\; \zeta =  -\frac{{G^{(a)}_\ell}''}{{F^{(a)}_\ell}'}.
\end{equation}
Then, (in the given approximation) the resonant partial absorption cross-section is reduced to the following universal, $q$- and $\ell$-independent form:
\begin{equation}\label{kappa}
    \kappa = \frac{\zeta}{(1+\zeta)^2+\xi^2};\;\; \zeta \geq 0.
\end{equation}
Thus, the line along $\xi$ (i.e., $\epsilon')$ axis has a typical symmetric Lorentzian shape, while the one along $\zeta$ (i.e., $\epsilon'')$ axis is strongly asymmetric, see also Fig.~\ref{F1}.

It is interesting to mention that at the resonant points of the absorption cross-sections given by Eqs.~(\ref{Re_e_l})-(\ref{Im_e_l}), the scattering cross-section is $Q^{(\ell)}_{sca}(\epsilon_\ell', \epsilon_\ell'') \simeq Q^{(\ell)}_{abs\; max}$, see Eq.~(\ref{Qmax}). At this point the partial extinction cross-section equals just double of $Q^{(\ell)}_{sca\; max}$.

The dimensional absorption cross-section which corresponds to Eq.~(\ref{Qmax}) $\sigma^{(\ell)}_{abs\;max} = \pi(2\ell+1)/2(n_m k_0)^2$ does not vanish at $R \rightarrow 0$, which looks confusing. The confusion is resolved if one notes that for the function $\sigma^{(\ell)}_{abs}(R,{\epsilon_\ell}',{\epsilon_\ell}'')$, the point $R = 0$ is singular; the function does not have any limit at this point and may take any value varying from 0 to $\sigma^{(\ell)}_{abs\;max}$ depending on the way to approach this point in $(R,{\epsilon_\ell}',{\epsilon_\ell}'')$ space. In particular, if one first fixes values of $\epsilon',\epsilon''$ and then tends $R$ to zero (which is the closest case to a possible experimental situation), $\sigma^{(\ell)}_{abs}(R,\epsilon',\epsilon'')$ vanishes, as it should be for a particle with zero radius, cf. the analogous properties of the resonant partial cross-sections of the anomalous scattering~\cite{MIT-Fano,MIT-Luk,MIT_JETP}.

Regarding off-resonant partial cross-sections, their values are given by the expression:
\begin{equation}\label{Qoff}
    Q^{(l)}_{abs} \simeq q^{2l-1}\epsilon''\frac{2(2l+1)(l+1)\ell^2}{[(\ell-l)(2l-1)!!]^2};\;\; l \neq \ell.
\end{equation}
Comparison with Eq.~({\ref{Qmax}) reveals that at small $q$ in the vicinity of the resonances, the contribution of the partial resonant cross-section to the net cross-section $Q_{abs}$ is overwhelming.

Let us discuss possible experimental observation of the anomalous absorption and related issues. In the preceding theoretical analysis three quantities $R,\;\epsilon'$ and $\epsilon''$ have been regarded as independent. In reality it is difficult (if possible) to tune $\epsilon'$ and $\epsilon''$ independently. Actual independent control parameters are $R$ and $\omega$. Then to maintain the resonance conditions, these quantities should satisfy the following set of equations:
\begin{eqnarray}
  \epsilon'(\omega, R) &=& \epsilon_\ell'(\omega, R), \nonumber \\
  \epsilon''(\omega, R) &=& \epsilon_\ell''(\omega, R), \nonumber
\end{eqnarray}
where the left-hand sides follows from the dispersion properties of the particle material (including spatial dispersion) and the right-hand sides are given by Eqs.~(\ref{Re_e_l}), (\ref{Im_e_l}), respectively. Solutions of the equations (if any) yield a unique discrete set of pairs $(\omega_\ell, R_\ell)$, where $R_\ell$ should satisfy the additional constraint following from the restriction $q \ll 1$. These conditions are very strict and the entire set of them is very difficult to fulfill in any real experiment. Thus, it seems the phenomenon discussed is just a ``virtual" effect.

Fortunately the situation is not so dramatic. There are several reasons for that. First, the resonance lines are narrow in the $(\epsilon', \epsilon'')$ plane, but this is not necessarily the case for the linewidths along the $\omega$ axis. The latter is determined by the dispersion properties of the particle materials. For most materials in the discussed region ($\epsilon'<-1$), the dispersion of $\epsilon$ is rather weak and the resonance conditions imposed on $\omega$ are not as strict as those for $\epsilon$. Second, while the decay of the absorption cross-section at $\epsilon' < \epsilon_\ell'$ is very sharp, it is rather slow at  $\epsilon' > \epsilon_\ell'$, see Fig.~(\ref{F1}b). Finally, one has to take into account sharpness of the $R$-dependence of the right-hand side of the resonance condition, see Eq.~(\ref{Im_e_l}). Due to all these facts, to observe the anomalous absorption it suffices to select material with small values of $\epsilon''$ for $\epsilon'$ lying in the region of possible resonances (in reality for $-2.5 \leq \epsilon' \leq -1.5$). Then, varying $R$ (or/and $\omega$), one inevitably passes through the vicinity of the resonance(s).

To illustrate this general reasoning, the absorption cross-section of aluminium nanoparticles is studied in detail. The model employed in the study is identical to the one discussed in Ref.~\cite{Libenson}: first, the dependence of $\epsilon$ on $\omega$ is taken from an empirical table~\cite{Palik}. Then, it is approximated by the Drude formula:
\begin{equation}\label{Drude}
    \epsilon = 1-\frac{\omega_p^2}{\omega^2+\gamma^2} + i\frac{\gamma\omega_p^2}{\omega(\omega^2+\gamma^2)}.
\end{equation}
To enhance the accuracy of the approximation, quantities $\omega_p$ and $\gamma$ are regarded as functions of $\omega$, calculated at every point in the table. This allows the identity of the initial table with the one resulting from Eq.~(\ref{Drude}). To take into account collisions of free electrons with the particle surface, the obtained $\gamma(\omega)$ for bulk aluminium is replaced by $\gamma_{eff} = \gamma(\omega) + v_F/R$, where $v_F = 10^8$~cm/s stands for the Fermi velocity of the free electrons. Next, polynomial interpolation between the table points is used to get a smooth $\epsilon(\omega)$, which is used to calculate the absorption cross-section derived from the exact Mie solution.
\begin{figure}[h]
  \hspace*{-2mm}
    \includegraphics[width=90mm]{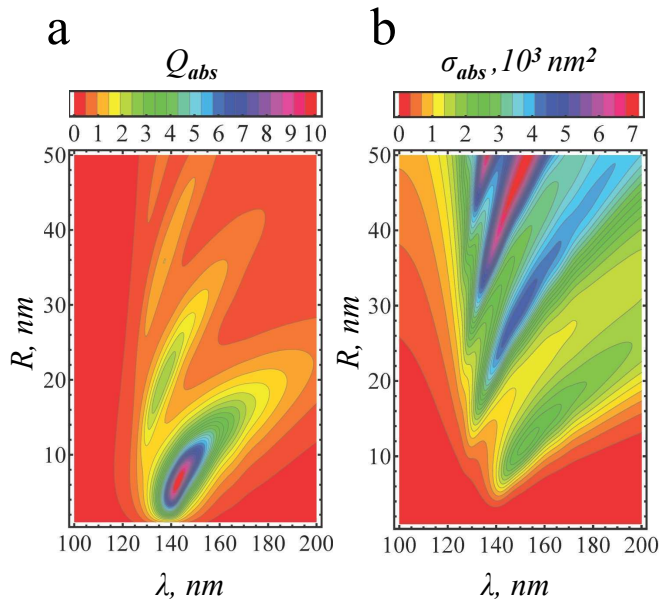}\\
  \vspace*{-3mm}
  \caption{(color online). Contour plots of the dimensionless (a) and dimensional (b) absorption cross-sections for an aluminium particle calculated based upon the exact Mie solution and the actual empirical dependence $\epsilon(\omega)$ for aluminium~\protect{\cite{Palik}}. Note that the localized maxima corresponding to different orders of the resonances for the cross-section normalized over $\pi R^2$ (a) become more extended and merge for the dimensional cross-section~(b).
  }\label{F2}\vspace*{-3mm}
\end{figure}
\begin{table}[h]
\caption{Approximate theory vs. exact Mie solution for $Al$ \label{Table}}
\begin{ruledtabular}
\begin{tabular}{c|c|c|c|c|c|c}
 $\ell$ & $\lambda_\ell,\!$ nm & $R_\ell,\!$ nm & $q$ &
\begin{tabular}{cc}
\multicolumn{2}{c}{$\epsilon_\ell'$} \\
\hline
Appr.& Exact
\end{tabular}
&
\begin{tabular}{cc}
\multicolumn{2}{c}{$\epsilon_\ell''$} \\
\hline
Appr. & Exact \end{tabular} &
$\Upsilon$ \\
 \hline
  1 & 143 & 6.45 & 0.283 & -2.16 \vline \hspace{1mm} -2.19 & 0.046 \vline \hspace{1mm} 0.216 & 0.617\\
  \hline
  2 & 136 & 19.27 & 0.890 & -1.67 \vline \hspace{1mm} -1.88 & 0.047 \vline \hspace{1mm} 0.159  &  0.901\\
 \end{tabular}
 \end{ruledtabular}
 \end{table}

Results of these calculations are presented in Fig.~\ref{F2}.  Quantitative comparison of the results with the developed approximate analytical theory is shown in Table~\ref{Table}. The resonant values $\lambda_\ell$ and $R_\ell$ correspond to the local maxima of Fig.~\ref{F1}a. The quantity $\epsilon_\ell'$ is calculated as a root of equation ${G^{(a)}}'_\ell(q,\epsilon')=0$, see Eq.~(\ref{Iaq}). The quality factor $\Upsilon$ equals the ratio of the net absorption cross-section (at a given $\lambda_\ell$ for the aluminium particle) to  $Q^{(\ell)}_{abs\;max}$, see Eq.~(\ref{Qmax}).

Note that the obtained  $\epsilon_\ell''$ for the aluminium particle are considerably \emph{larger\/} than those which, according to Eq.~(\ref{Im_e_l}), should correspond to the resonances; nevertheless well-pronounced maxima of the dipole, quadrupole and even octupole resonances are seen on Fig.~\ref{F1} straightforwardly. The absolute maximum of the relative absorption cross-section, Max\{$Q_{abs}$\} $\approx 11.516$, is achieved at $\lambda \approx 143$ nm; $R \approx 6.45$ nm and corresponds to the dipole resonance ($\ell = 1$).

The presented analysis has thusly revealed a small particle made of weakly dissipating materials has the absorption cross-section which may exceed its geometric cross-section in order of magnitude, and even more. Though the anomalous absorption and the anomalous scattering correspond to opposite limits (the maximal and vanishing absorption, respectively), they have much in common, namely both are associated with giant amplification of incident light inside the scattering particle in the vicinity of the plasmon (polariton) resonances accompanied by interplay of the radiative and dissipative dampings, both cannot be described within the framework of the dipole approximation and require analysis of the exact Mie solution, and both may exhibit similar unusual properties (the inverted hierarchy of the resonances, singular dependencies of cross-sections on the particle size, etc.). The calculations for aluminium particles indicate that the anomalous absorption should be an experimentally observable phenomenon. A future endeavor will be to carry out the corresponding experimental study, which hopefully may be motivated by this Letter.



\begin{thebibliography}{99}
\bibitem{n1}It does not effect convergence of the multipole expansion because (i) for a spherical particle each resonance occurs at its own frequency; (ii) the inverted hierarchy lasts up to a certain finite order of the resonances, then the normal hierarchy is restored~\cite{MIT_JETP}.
\end{thebibliography}
\end{document}